\documentclass[a4paper,11pt]{article}
\usepackage{pos}
\usepackage{amsmath}
\usepackage{slashed}
\usepackage{multirow}
\usepackage{subfigure}
\usepackage{bbm}

\title{Portable Lattice QCD implementation based on OpenCL}

\author*[a]{Piyush Kumar}
\author[a]{Szabolcs Borsanyi}
\author[a]{Jana N. Guenther}
\author[a]{Chik Him Wong}

\affiliation[a]{Bergische Universität Wuppertal,\\
  Gaussstr. 20, Wuppertal, Germany}

\emailAdd{kumar@uni-wuppertal.de}
\emailAdd{borsanyi@uni-wuppertal.de}
\emailAdd{jguenther@uni-wuppertal.de}
\emailAdd{cwong@uni-wuppertal.de}

\abstract{The presence of GPU from different vendors demands the Lattice QCD codes to support multiple architectures. To this end, Open Computing Language (OpenCL) is one of the viable frameworks for writing a portable code. It is of interest to find out how the OpenCL implementation performs as compared to the code based on a dedicated programming interface such as CUDA for Nvidia GPUs. We have developed an OpenCL backend for our already existing code of the Wuppertal-Budapest collaboration. In this contribution, we show benchmarks of the most time consuming part of the numerical simulation, namely, the inversion of the Dirac operator. We present the code performance on the JUWELS and LUMI Supercomputers based on Nvidia and AMD graphics cards, respectively, and compare with the CUDA backend implementation.}

\FullConference{The 41st International Symposium on Lattice Field Theory (LATTICE2024)\\
 28 July - 3 August 2024\\
Liverpool, UK\\}

\begin{document}
\maketitle

\section{Introduction}

One of the major challenges in investigating the physics of quarks and gluons using lattice QCD is the high computational cost, which grows significantly as the lattice size increases. A key step in the simulation using Hybrid-Monte-Carlo (HMC) algorithm is the inversion of the large, sparse fermionic matrix, $D$ of size $N \times N$, where $N = 3 \times N_\sigma^3 \times N_\tau$ with staggered fermions on an $N_\sigma^3 \times N_\tau$ lattice. The inversion using iterative methods such as Conjugate Gradient (CG) requires repeatedly applying the fermionic matrix on a pseudo-fermion, $\slashed{D}\phi$ which is computationally expensive and is the most time critical part of the simulation.\\
The high computational demand and the inherent parallelism in lattice QCD simulations have popularized the utilization of GPUs. This has been crucial in attaining the performance required to make the physical calculations computationally feasible. Since the first Lattice QCD implementation on GPUs using Open Graphics Library (OpenGL) in 2006 \cite{Egri:2006zm}, offloading the expensive parts of the simulation to GPUs has become a standard practice. This led to the development of optimized software packages and frameworks for lattice QCD simulations such as QUDA \cite{Clark:2009wm}, Grid \cite{Boyle:2015tjk} and SimulateQCD \cite{HotQCD:2023ghu} which provide multi-GPU support for the accelerated simulations. Additionally, the presence of GPU from different vendors demands the Lattice QCD codes to support multiple architectures. To this end, Open Computing Language (OpenCL) \cite{5457293} and SYCL \cite{SYCL2020} are the viable frameworks to write a portable code. We have added an OpenCL backend (see also \cite{Bach:2012iw}) to our already existing code. In this work, we show the benchmarks of the $\slashed{D}$ operator in the staggered formalism and its inversion using CG on a single GPU.

\section{OpenCL}
\begin{figure}[h!]
    \centering
    \includegraphics[width=0.7\linewidth]{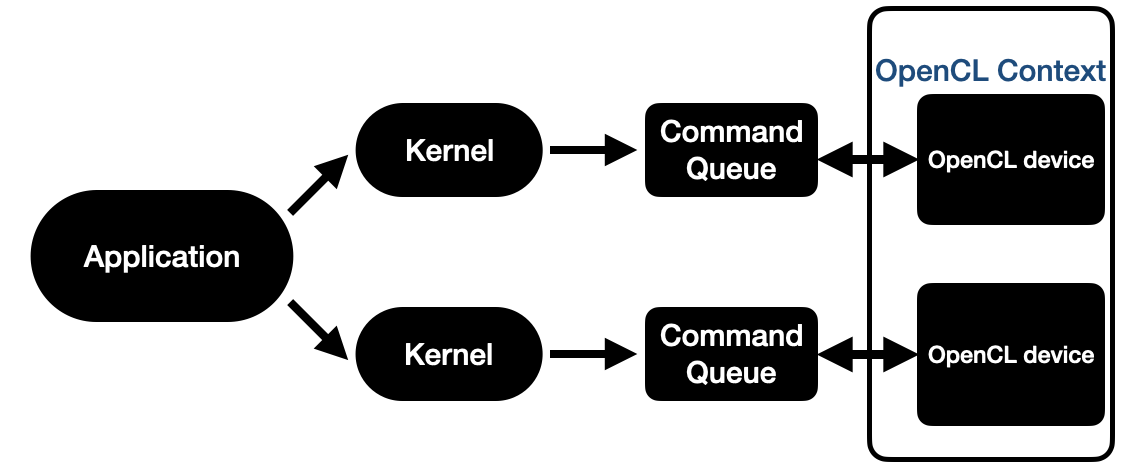}
    \caption{Schematic of OpenCL kernel execution model}
    \label{opencl_exec}
\end{figure}

OpenCL is an open standard for writing parallel programs to run across heterogeneous platforms and architectures. The OpenCL API provides a standard interface for developing kernels, which ensures code portability across different hardware architectures, unlike vendor-specific options, such as CUDA and HIP. The execution model consists of a host program and kernels. The host code uses the OpenCL API to query and select compute devices, associate a context for the devices and, within the context, create a command queue for each device to manage the workload. The lower level of the execution model consists of "Kernel", i.e. device code which is executed on the processing elements within the compute device. As shown in Fig. \ref{opencl_exec}, when the kernels are called by the application, they are enqueued to the command queue of the device along with the work size and work group decomposition. The work size defines the total number of work items (threads) to be executed, while the work group size specifies how these work items are grouped for execution on compute units. Once the dependencies of the kernel-instance are met, i.e., completion of previously enqueued synchronization commands, OpenCL runtime launches the kernel-instance across the available compute units of the device for execution.
OpenCL offers the runtime compilation and building of the source code for device kernels. This offers several benefits, such as greater portability, just-in-time hardware optimizations and the flexibility of dynamically generating or modifying kernels at runtime based on input data and application requirements.
\section{Code Design}
\begin{figure}[h!]
    \centering
    \includegraphics[width=0.65\linewidth]{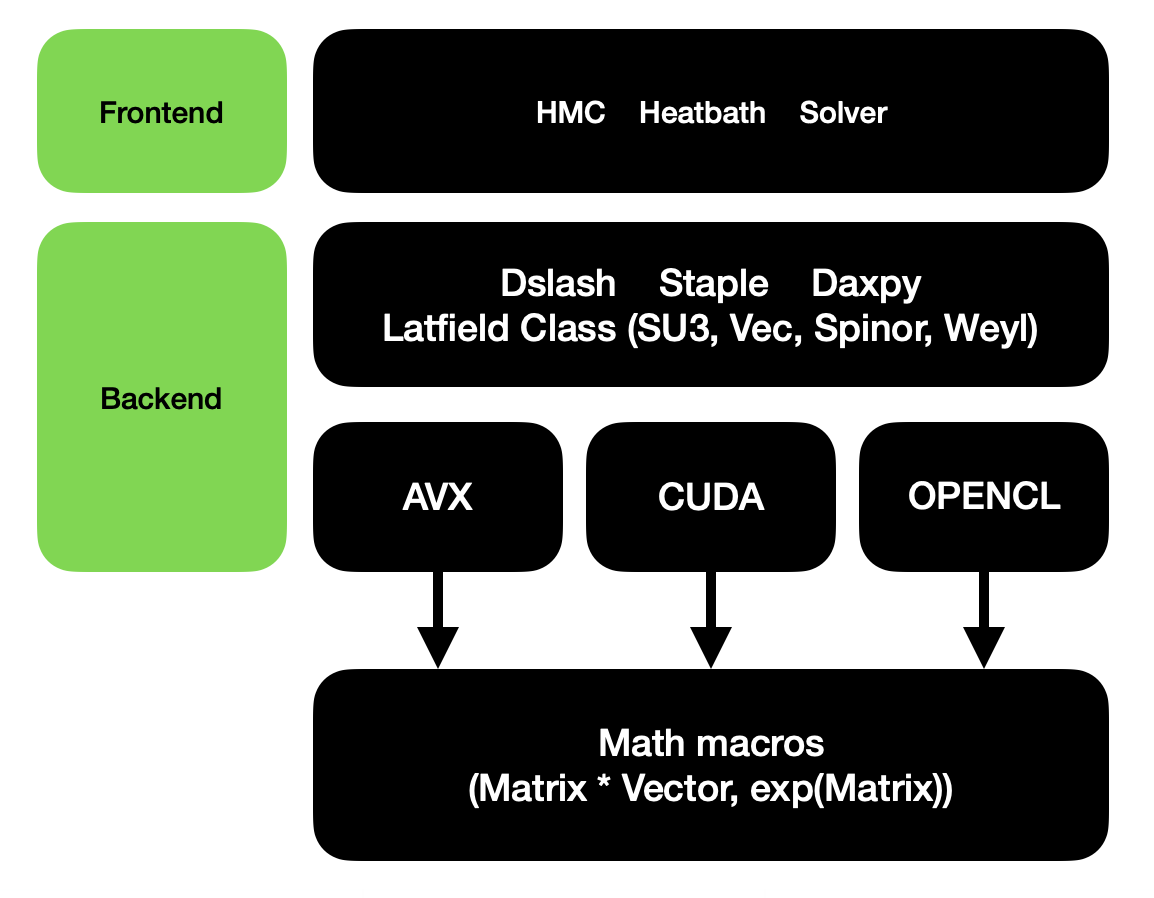}
    \caption{Diagram illustrating the code hierarchy by providing some example classes}
    \label{Newcode_hierarchy}
\end{figure}

We show the organizational levels of the code in Fig. \ref{Newcode_hierarchy}. At the top layer, we have the tasks or algorithms, such as HMC, parallel tempering, density of states and iterative solvers. The backends provide the implementations of the Dirac operator, the gauge action and its derivatives, etc. The backend contains templated classes for the lattice fields ("Latfield" class) such as four dimensional link fields, spinor fields, etc. Each backend optimizes the memory layout and management of field objects for the target architecture. In the OpenCL backend, we store Latfield objects in the Structure of Array (SoA) format, with sites organized such that all even sites are listed first, followed by all odd sites. Both CPU and GPU based implementations rely on a shared library of macros defined for the sitewise mathematical computations, which sits at the lowest level of the hierarchy.

\section{Benchmarks}
\subsection{System specifications}
We tested our code on LUMI \cite{lumi_webpage} and JUWELS \cite{juwels_webpage} supercomputers across various lattice sizes up to $64^4$. The GPU partition of LUMI is configured with 4 AMD MI250X cards per node. Each MI250X consists of two Graphic Compute Dies (GCD), connected via Infinity Fabric with 200GB/s bi-directional bandwidth. Each GCD of AMD MI250x card has a High-Bandwidth Memory (HBM) bandwidth of 1.64 TB/s and a theoretical peak performance of $28.16$ TFlop/s. The Booster module, on the other hand, has 4 NVIDIA A100 Tensor core GPUs per node. A100s have an HBM bandwidth of $1.55$ TB/s and a peak performance of $19.5$ TFlop/s and $9.7$ TFlop/s for single (FP32) and double (FP64) precision respectively. For the benchmarking, we evaluated the performance of our code on a single MI250X GCD and a single A100 GPU.

\subsection{Dslash Operator and its inversion}
We consider the staggered $\slashed{D}$ operator on a colour vector field $\phi_x$ which reads:
\begin{equation}
    \slashed{D} \phi_x = \sum_{\mu=0}^{3} \left[ \eta_{\mu}(x) ~ U_{x,\mu} \phi_{x + \hat{\mu}} - \eta_{\mu}(x-\hat{\mu}) ~ U_{x-\hat{\mu},\mu}^\dagger \phi_{x-\hat{\mu}}\right]
    \label{staggered}
\end{equation}
where the $U_\mu$ are the $SU(3)$ link variables and $\eta_\mu(x) = \pm 1$ is the Kogut-Susskind phase factor.
\begin{figure}[h!]
            \centering
            \subfigure{
               \centering\includegraphics[width=0.325\paperwidth]{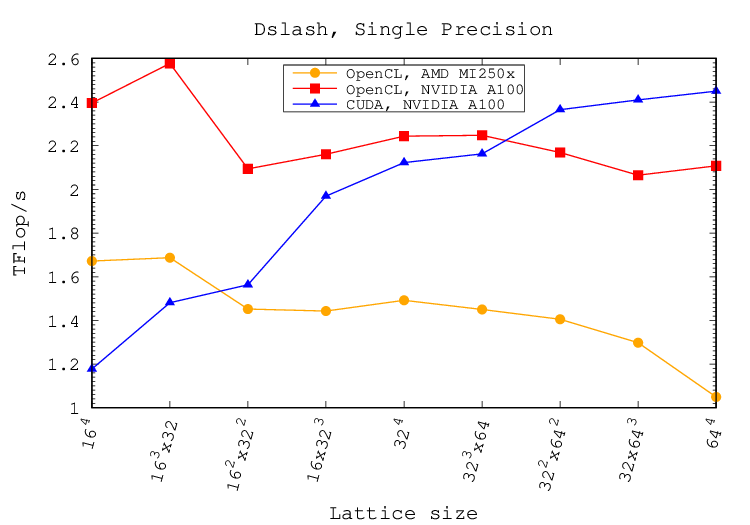}
            }
            \subfigure{
                \centering\includegraphics[width=0.325\paperwidth]{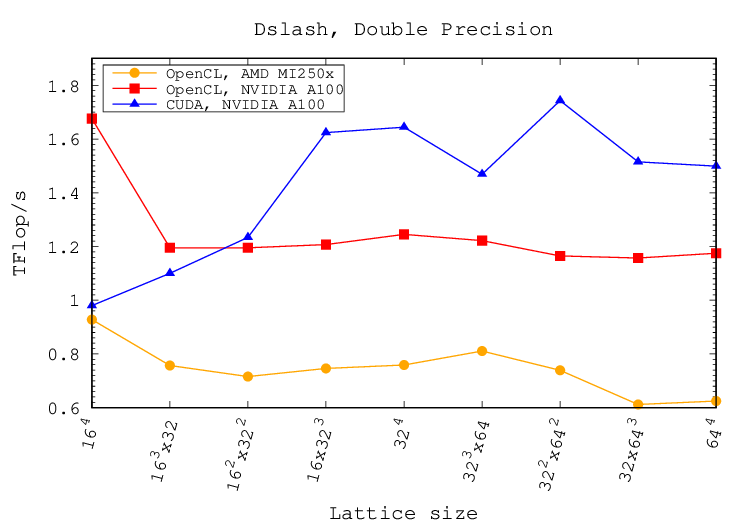}
            }
            \vspace{0.7cm}
            \subfigure{
                \centering\includegraphics[width=0.325\paperwidth]{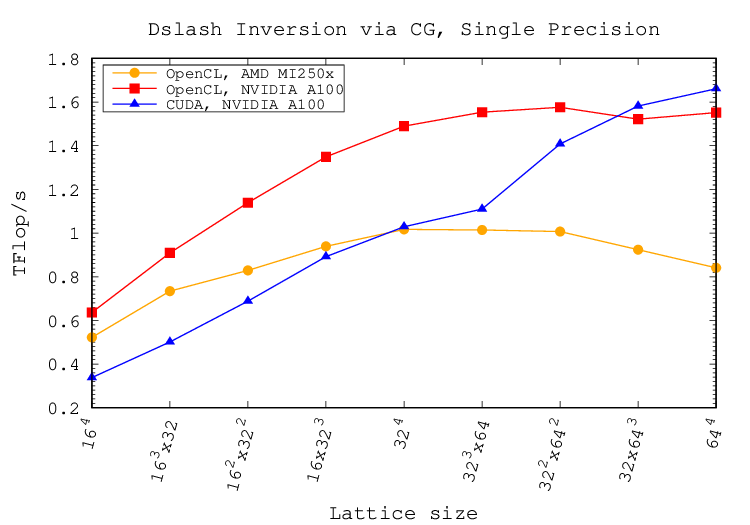}
            }
           \subfigure{
                \centering\includegraphics[width=0.325\paperwidth]{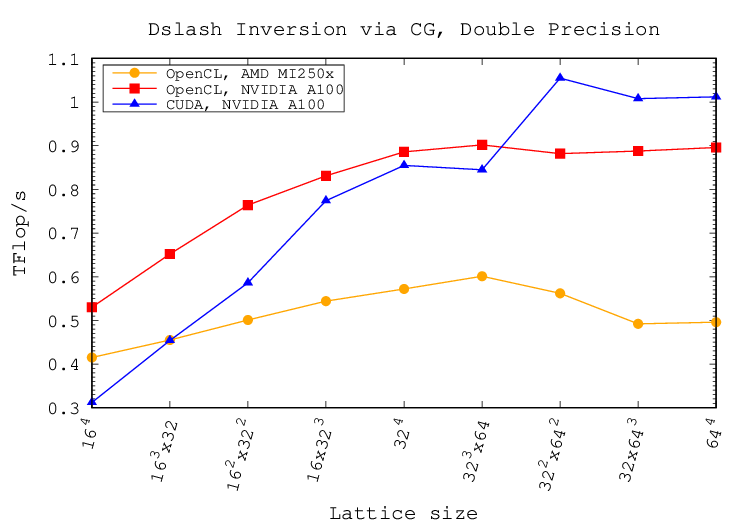}
            }
         \caption{Comparison of benchmarks for $\slashed{D}$ and its inversion via CG in single and double precision.}
         \label{fig:benchmarks}
\end{figure}
Fig. \ref{fig:benchmarks} showcases the benchmarks of our OpenCL implementation, comparing it with the CUDA counterpart for the Dslash kernel and CG solver. The performance of OpenCL implementation for Dslash peaks around $16^4$ lattice size and stays mostly constant after that. However, for a single gcd of MI250x, we observe degradation in performance for sizes beyond $32^2 \times 64^2$. The CUDA backend uses a different memory layout which is better suited for larger lattice sizes, thus, giving higher performance than OpenCL for such cases. The inversion of Dirac operator requires other vector routines apart from Dslash, such as dot product involving global reductions which have higher latencies. These effects are more pronounced for smaller lattice sizes which explains the large gap between the Dslash and CG benchmarks. For both Dslash and CG, the performance on one A100 is roughly $1.5$ times that of a single gcd of MI250x. In table \ref{table_of_bw}, we present the achieved bandwidth (in GB/s) for the double precision CG update routine on a $48^4$ lattice. For the A100 GPU, we were able to achieve approximately 85\% of the peak bandwidth, similar to the CUDA backend. In contrast, for the MI250X, this number is relatively lower at around 75\% of the theoretical maximum bandwidth.
\begin{table}[h!]
        \centering
        \begin{tabular}{ c|c|c }
        \hline
        \textbf{Implementation}  & \textbf{Sustained bandwidth} & \textbf{Peak Bandwidth} \\
        \hline \hline
        OpenCL on MI250X  & $1134$  & 1638\\
        \hline
        OpenCL on A100 & $1351$  & 1555\\
        \hline
        CUDA on A100 & $1317$  & 1555 \\
        \hline
        \end{tabular}
            \caption{Bandwidth (GB/s) for CG update routine on a $48^4$ lattice.}
            \label{table_of_bw}
    \end{table}

\subsection{Dslash Roofline}
\begin{figure}[h!]
    \centering
    \includegraphics[width=0.65\linewidth]{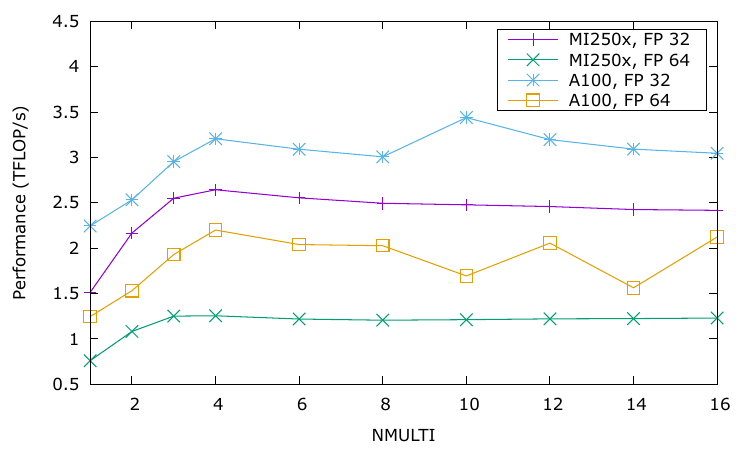}
    \caption{Performance of Dslash with varying number of vectors on one A100 and a single gcd of MI250x on a $32^4$ lattice}
    \label{nmulti_plot}
\end{figure}

The roofline plot gives a bound on attainable performance based on the device capabilities such as peak performance and memory bandwidths. The peak performance (Flop/s) of a kernel is defined as follows:
\begin{equation}
\text{Attainable performance} = \min \left\{
\begin{array}{c}
\text{peak performance} \\
\text{arithmetic intensity} ~ \times ~ \text{max bandwidth} 
\end{array}
\right\}
\end{equation}
We need to compute the arithmetic intensity (Flops/byte) of the staggered Dslash kernel to which we now turn. From eq. (\ref{staggered}), for each direction $\mu$, the kernel performs two complex $3 \times 3$ matrix vector products and one complex 3-vector addition. This amounts to a total flop count of $570$ at each lattice site. We use gauge fixing to transform the temporal link variables to $\mathbbm{1}_{3\times3}$ for all the sites with the temporal coordinate, $t \neq 0$. For each link variable, we store only the first two rows, i.e. 12 floats. The third row is reconstructed on-the-fly when needed, using the properties of SU(3) matrices. So, in terms of memory, the kernel reads 6 $SU(3)$ matrices, 8 complex 3-vectors and writes 1 complex 3-vector giving a total float count of 126 per site. The arithmetic intensity of the kernel is $\sim 0.57$ Flops/Byte for FP64 and $\sim 1.13$ Flops/Byte for FP32. Since the performance is bounded by the device bandwidth, we can increase the arithmetic intensity and potentially improve the performance by applying $\slashed{D}$ on multiple vectors at once. This allows each link variable to be read once and then reused for multiple vectors, thereby decreasing the total number of memory accesses required. For example, the arithmetic intensity as a function of the number of vectors, $n$ is given by $n\cdot570/(8\cdot(6\cdot12 + n(9\cdot6)))$ for FP64. In the code, we set the parameter $n=$ NMULTI $>1$ to get the benchmarks with varying the number of vectors for $32^4$ lattice, as summarized in Table \ref{nmulti_performance_data} and Fig. \ref{nmulti_plot}. For both A100 and MI250x, we see consistent improvements in the performance up to NMULTI=4, after which, it saturates.

\begin{figure}[h!]
            \centering
            \subfigure{
               \centering\includegraphics[width=0.325\paperwidth]{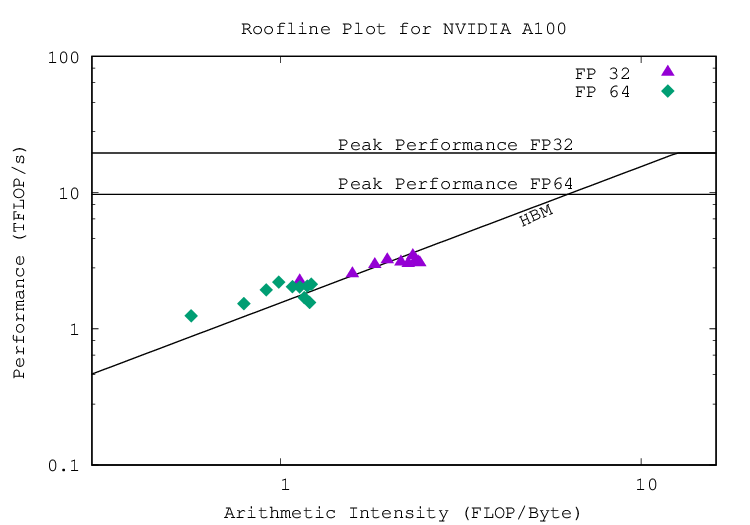}
            }
            \subfigure{
                \centering\includegraphics[width=0.325\paperwidth]{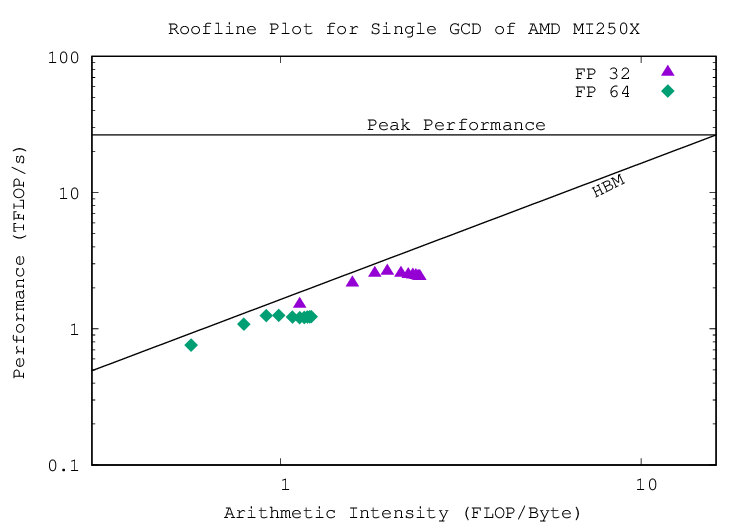}
            }
         \caption{Roofline plots for staggered Dslash operator with varying number of vectors on a $32^4$ lattice on one A100 and a single gcd of MI250x.}
         \label{fig:roofline}
\end{figure}

\begin{table}[h]
\centering
\begin{tabular}{|c|c|c|c|c|}
\hline
NMULTI & \multicolumn{2}{c|}{A100} & \multicolumn{2}{c|}{MI250x} \\
\cline{2-5}
 & FP32 & FP64 & FP32 & FP64 \\
\hline
1  & 2.244 & 1.245 & 1.512 & 0.759  \\
2  & 2.532 & 1.530 & 2.166 & 1.081 \\
3  & 2.953 & 1.929 & 2.550 & 1.250 \\
4  & 3.205 & 2.199 & 2.643 & 1.254 \\
6  & 3.090 & 2.039 & 2.554 & 1.218 \\
8  & 3.006 & 2.026 & 2.494 & 1.205 \\
10 & 3.439 & 1.693 & 2.477 & 1.211 \\
12 & 3.196 & 2.055 & 2.458 & 1.220 \\
14 & 3.091 & 1.562 & 2.425 & 1.223 \\
16 & 3.045 & 2.122 & 2.416 & 1.229 \\
\hline
\end{tabular}
\caption{Performance of Dslash (in TFLOP/s) with varying number of vectors on a $32^4$ lattice on one A100 and a single gcd of MI250x.}
\label{nmulti_performance_data}
\end{table}
We show the roofline plots for Dslash operator on a $32^4$ lattice in Fig. \ref{fig:roofline}. For A100, we consistently stay above the roofline which suggests the effective usage of caching. For MI250x on the other hand, since the performance stagnates for NMULTI $>3$, we see the drift away from the roofline.  

\section{Conclusions and Outlook}
Our portable OpenCL implementation for a single GPU achieves performance on par with CUDA across a range of lattice sizes. With the current version, the use case of the OpenCL
implementation includes lattice QCD simulations at finite temperature which is carried out on moderate lattice sizes that fit in a single GPU. The OpenCL backend is already being utilized in production runs, particularly for the study of QCD equation of state using density of states method. We are presently incorporating multi-GPU support to our portable implementation. It is not straightforward to do so with OpenCL, since this API lacks the direct GPU-GPU communication capabilities. As an alternative we are also working to develop a SYCL backend which will overcome this shortcoming with GPU aware MPI, while maintaining the portability features of our OpenCL implementation.

\begin{acknowledgments}
This work is supported by the MKW NRW under the funding code NW21-024-A.
Further funding was received from the DFG under the Project
No. 496127839.
The authors gratefully acknowledge the Gauss Centre for
Supercomputing e.V. (\url{www.gauss-centre.eu}) for funding
this project by providing computing time on the GCS
Supercomputer Juwels-Booster at Juelich Supercomputer
Centre.
We acknowledge the EuroHPC Joint Undertaking for awarding this project access to the EuroHPC supercomputer LUMI, hosted by CSC (Finland) and the LUMI consortium through a EuroHPC Extreme Access call.
\end{acknowledgments}

\bibliography{ref.bib}

\providecommand{\href}[2]{#2}\begingroup\raggedright\begin{thebibliography}{1}

\bibitem{Egri:2006zm}
G.I.~Egri, Z.~Fodor, C.~Hoelbling, S.D.~Katz, D.~Nogradi and K.K.~Szabo,
  \emph{{Lattice QCD as a video game}},
  \href{https://doi.org/10.1016/j.cpc.2007.06.005}{\emph{Comput. Phys. Commun.}
  {\bfseries 177} (2007) 631}
  [\href{https://arxiv.org/abs/hep-lat/0611022}{{\ttfamily hep-lat/0611022}}].

\bibitem{Clark:2009wm}
{\scshape QUDA} collaboration, \emph{{Solving Lattice QCD systems of equations
  using mixed precision solvers on GPUs}},
  \href{https://doi.org/10.1016/j.cpc.2010.05.002}{\emph{Comput. Phys. Commun.}
  {\bfseries 181} (2010) 1517}
  [\href{https://arxiv.org/abs/0911.3191}{{\ttfamily 0911.3191}}].

\bibitem{Boyle:2015tjk}
P.~Boyle, A.~Yamaguchi, G.~Cossu and A.~Portelli, \emph{{Grid: A next
  generation data parallel C++ QCD library}},
  \href{https://arxiv.org/abs/1512.03487}{{\ttfamily 1512.03487}}.

\bibitem{HotQCD:2023ghu}
{\scshape HotQCD} collaboration, \emph{{SIMULATeQCD: A simple multi-GPU lattice
  code for QCD calculations}},
  \href{https://doi.org/10.1016/j.cpc.2024.109164}{\emph{Comput. Phys. Commun.}
  {\bfseries 300} (2024) 109164}
  [\href{https://arxiv.org/abs/2306.01098}{{\ttfamily 2306.01098}}].

\bibitem{5457293}
J.E.~Stone, D.~Gohara and G.~Shi, \emph{Opencl: A parallel programming standard
  for heterogeneous computing systems},
  \href{https://doi.org/10.1109/MCSE.2010.69}{\emph{Computing in Science \&
  Engineering} {\bfseries 12} (2010) 66}.

\bibitem{SYCL2020}
{Khronos Group}, ``Sycl 2020 specification.''
  \url{https://www.khronos.org/registry/SYCL/specs/sycl-2020/pdf/sycl-2020.pdf},
  2021.

\bibitem{Bach:2012iw}
M.~Bach, V.~Lindenstruth, O.~Philipsen and C.~Pinke, \emph{{Lattice QCD based
  on OpenCL}}, \href{https://doi.org/10.1016/j.cpc.2013.03.020}{\emph{Comput.
  Phys. Commun.} {\bfseries 184} (2013) 2042}
  [\href{https://arxiv.org/abs/1209.5942}{{\ttfamily 1209.5942}}].

\bibitem{lumi_webpage}
{LUMI}, ``{LUMI Supercomputer}.'' \url{https://www.lumi-supercomputer.eu/},
  2025.

\bibitem{juwels_webpage}
{Jülich Wizard for European Leadership Science (JUWELS)}, ``{JUWELS
  Supercomputer}.''
  \url{https://www.fz-juelich.de/en/ias/jsc/systems/supercomputers/juwels},
  2025.

\end{thebibliography}\endgroup
\bibliographystyle{JHEP.bst}

\end{document}